\documentclass[twocolumn]{article}

\usepackage[utf8]{inputenc}
\usepackage[margin=1.25in]{geometry}
\usepackage{authblk}
\usepackage{setspace}
\usepackage{graphicx}
\usepackage{subcaption}
\graphicspath{{./figures/}}
\usepackage{subcaption}
\usepackage{amsmath}
\usepackage{booktabs}
\usepackage{hyperref}
\usepackage{xcolor}
\usepackage{dblfloatfix}
\usepackage{siunitx}

\usepackage[switch]{lineno}  
\usepackage[numbers,sort&compress]{natbib}

\title{Investigation of the in-pixel response of the Mupix11 monolithic pixel sensor using a microfocus X-ray beam at Diamond Light Source}

\author[1*]{A.S. Rotelli}
\author[3]{H. Augustin}
\author[1]{D. Bortoletto}
\author[1]{A. Brooks}
\author[2]{M. Grimes}
\author[1]{A.J.A. Knight}
\author[2]{M.S. Köppel}
\author[1]{A.E. McDougall}
\author[1]{R. Plackett}
\author[1]{D M S Sultan}
\author[3]{L. Vigani}
\author[1,4]{S. Wood}

\affil[1]{Department of Physics, University of Oxford, United Kingdom.}
\affil[2]{ETH Zurich, Switzerland.}
\affil[3]{Physikalisches Institut, University of Heidelberg, Germany.}
\affil[4]{Diamond Light Source Ltd., United Kingdom.}
\affil[*]{Address correspondence to: aliki.rotelli@physics.ox.ac.uk}

\date{}

\begin{document}
\twocolumn[
\begin{@twocolumnfalse}

\maketitle

\begin{abstract}
MuPix11 is a High-Voltage Monolithic Active Pixel Sensor (HV-MAPS) developed for the tracking system of the Mu3e experiment. The in-pixel
photon response of a MuPix11 sensor thinned to \SI{70}{\micro\meter} was measured using an \SI{8}{keV} X-ray beam with a \SI{3}{\micro\meter} spot size at the B16 beamline at Diamond Light Source, emulating the passage of a minimum ionising particle (MIP). At nominal operating voltage and threshold, high-resolution scans across the pixel matrix show the detector response to be uniform. In the absence of reverse bias (\SI{0}{V}), the relative sub-pixel response is location-dependent as a reduced detection rate is observed at pixel boundaries. 

\end{abstract}
\vspace{1em}
\end{@twocolumnfalse}
]

\section{Introduction}
The MuPix11, a member of the MuPix family of chips \cite{MuPix11}, is a High-Voltage Monolithic Active Pixel Sensor (HV-MAPS) \cite{MAPS} that has been developed for the Mu3e experiment \cite{TDR}.
This paper reports on in-pixel photon response and provides an estimate for the the relative X-ray detection efficiency of MuPix11.  From these measurements, the depletion state at different positions within the pixel structure can be inferred.  The measurements were performed using a micro-focus X-ray beam at the B16 beamline at Diamond Light Source, the UK’s national synchrotron facility, which provides highly configurable optics. \cite{Diamond}

\subsection{HV-MAPS}
\label{subsec:HVMAPS}
\begin{figure}
    \centering
    \includegraphics[width=\linewidth]{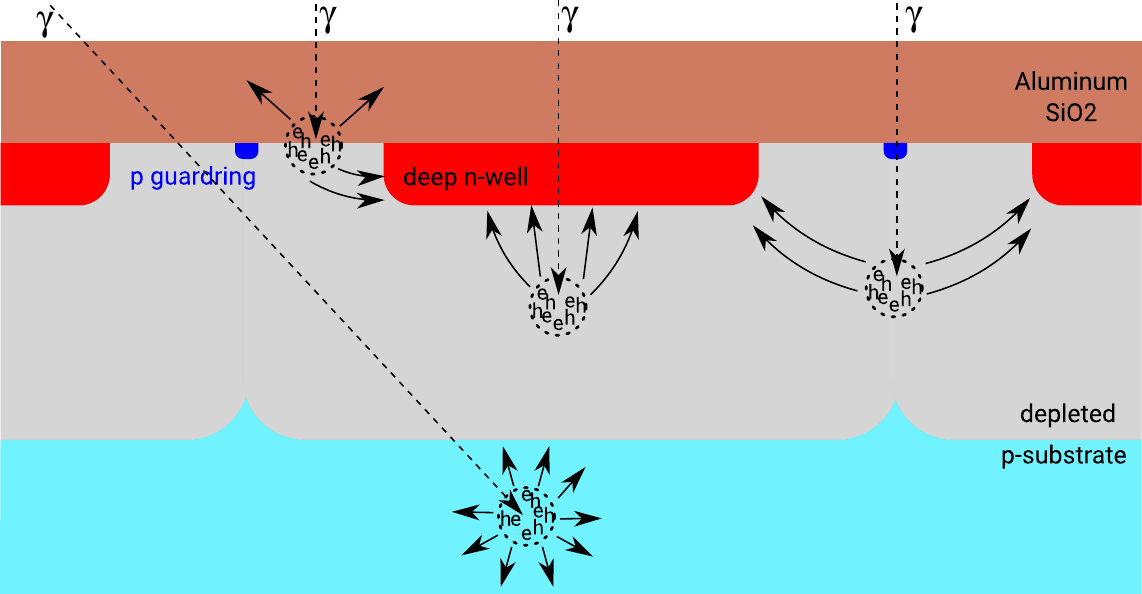}
    \caption{Schematic of the HV-MAPS pixel structure implemented in MuPix11 \cite{Immig}.}
    \label{fig:HVMAPS}
\end{figure}
HV-MAPS are a class of depleted silicon pixel detectors in which the pixel readout electronics and the depleted silicon sensing volume are integrated into the same silicon device. Compared to hybrid detectors they allow for thinner sensors, reduced multiple scattering length, and lower production cost.
In HV-MAPS, a deep n-well collecting diode is embedded in a p-type substrate, forming a p-n junction. As shown in Fig.~\ref{fig:HVMAPS}, in the MuPix11 design the n-well is located in the centre of the pixel. With increasing bias voltage, a depletion region forms around and below the n-well. However, even at \SI{0}{\volt} bias, a small built-in potential arising from the p-n junction enables some signal collection. For the MuPix11 this intrinsic voltage is \SIrange{0.61}{0.67}{V} \cite{Sze}. Additionally the n-well is biased to a voltage close to positive supply voltage, totalling at more than \SI{2}{\volt} depletion voltage even for \SI{0}{\volt} reverse bias, already generating a sizable depleted volume.

\subsection{MuPix11 in the Context of the Mu3e Experiment}
The MuPix11 sensor is designed for the tracking system of the Mu3e experiment \cite{TDR} by the Karlsruher Institut für Technologie (KIT) and collaborators. It is part of a family of ASICs that also includes the ATLASPix \cite{atlaspix}, and TelePix \cite{telepix} detectors. These devices share the characteristic that the deep n-well, which shields the integrated electronics from the sensor volume, also serves as a large primary charge collection electrode. This results in a large fill factor and facilitates uniform depletion of the sensor volume. MuPix11 features a pixel pitch of \SI{80}{\micro\meter} and provides an \SI{8}{ns} timestamp in an event-driven readout mode. The final Outer Pixel Detector will comprise 2,736 MuPix11 sensors, corresponding to approximately one square meter of silicon.

The tracking of low energy (\SI{10}{MeV} -- \SI{53}{MeV}) electrons, the primary task of the detector, requires an ultra-low-mass design to reduce multiple scattering. This has led Mu3e to adopt a four-layer barrel geometry design that provides no tracking redundancy. The $\mu \rightarrow eee$ signal requires the successful reconstruction of all three decay electrons, which must each produce hits in at least four detector layers. The overall signal efficiency strongly depends on the single-hit efficiency, $\epsilon$. To first approximation, the event efficiency therefore scales as $ \epsilon^{12}$, emphasising the importance of achieving near-unity hit efficiency.

\section{Experimental Procedure}
\label{sect:procedure}
The aim of this study is to measure a two-dimensional in-pixel response  map of MuPix11 pixels. The B16 beamline's micro-focused beam, produced using a Compound Refractive Lens (CRL) mirror system, was focused to a spot size of \SI{3}{\micro\meter} by \SI{1.8}{\micro\meter}. The photon energy was set to be 8~keV, as discussed in Section~\ref{subsec:beam}. 

The MuPix11 sensor was mounted on a precision motion stage and scanned through the beam in a raster pattern. Several thousand photons were recorded by the MuPix11 at each position. This procedure yields a map of the pixel response, i.e. the number of hits recorded as a function of beam position within the pixel. Measurements were performed at various sensor reverse-bias voltages and threshold settings to study the sensor performance.

\subsection{Beam Energy}
\label{subsec:beam}
For a \SI{70}{\micro\meter}-thick MuPix11, the active silicon thickness is estimated to be approximately \SI{50}{\micro\meter}. From the Landau-Vavilov\cite{PDG} theory, the most probable energy loss of a Minimum Ionising Particle (MIP) in \SI{50}{\micro\meter} of silicon \cite{riegler} corresponds to the creation of approximately 3,160 electron-hole pairs. Since the average energy required to create a single electron-hole pair in silicon is \SI{3.6}{eV} \cite{lutz}, this is equivalent to an X-ray energy of \SI{11.4}{keV}. 

However, the photoabsorption probability decreases with increasing photon energy. At \SI{11.4}{keV}, only 23\% of photons are absorbed in \SI{50}{\micro\meter} of silicon \cite{nist}, whereas at 8~keV the absorption probability increases to 52\%. Therefore, an X-ray energy of 8~keV was selected as a compromise between charge deposition and absorption probability.

An 8~keV photon produces approximately 2,220 electron–hole pairs, corresponding to about 70\% of the charge deposited by a MIP, which is sufficient for efficient collection by the MuPix11 charge-sensitive amplifier. In contrast to charged particles, which deposit energy continuously along their trajectory, 8~keV photons are predominantly absorbed via the photoelectric effect, resulting in a highly localised charge cloud. This introduces a small systematic difference with respect to the response to traversing charged particles.

\subsection{Setup}
\begin{figure}
    \centering
    \includegraphics[width=1\linewidth]{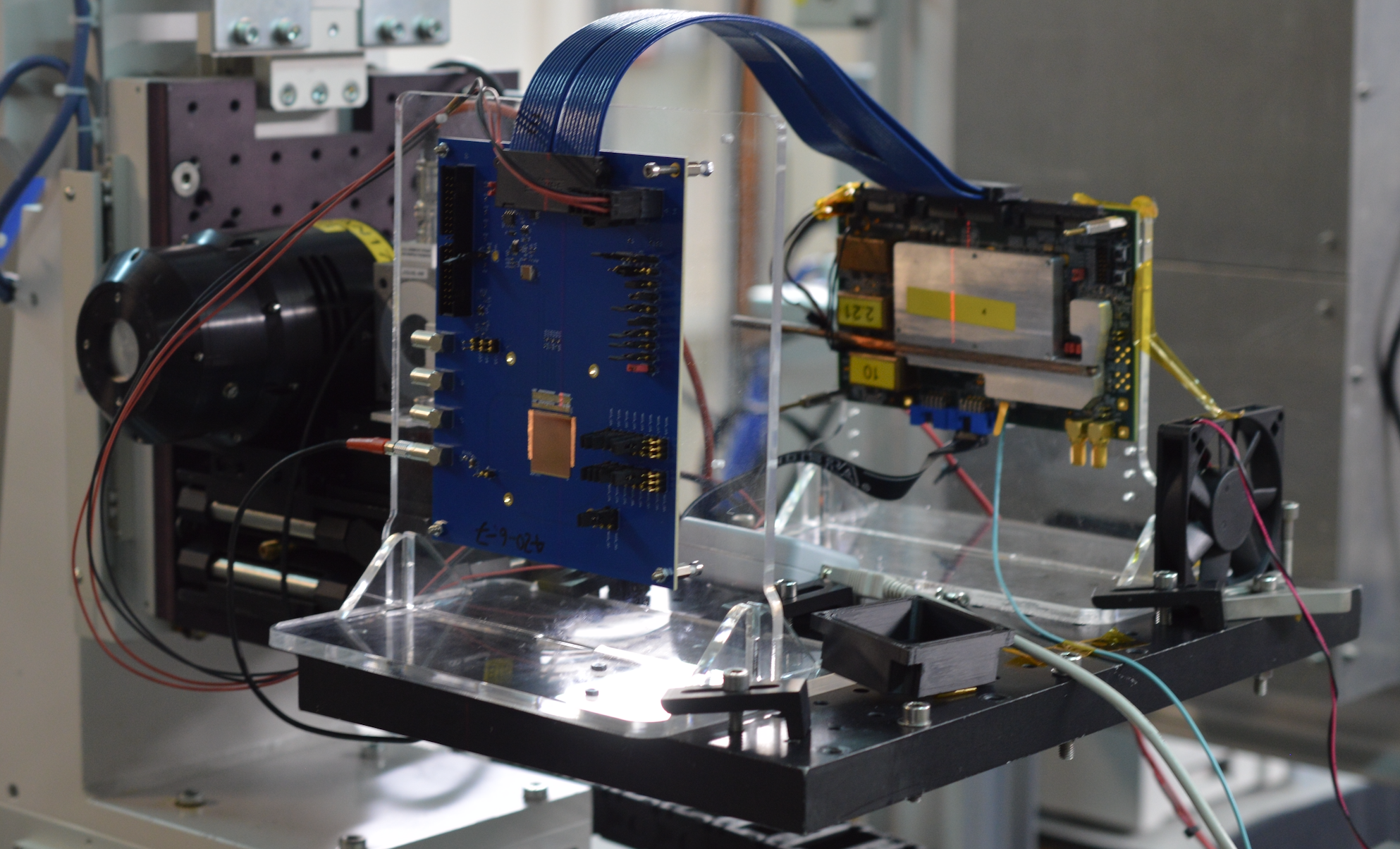}
    \caption{SCC and FEB mounted on the motion stage. The SCC provides connections for low- and high-voltage supplies and to the FEB.}
    \label{fig:setup}
\end{figure}

\begin{figure}
    \centering
    \includegraphics[width=1\linewidth]{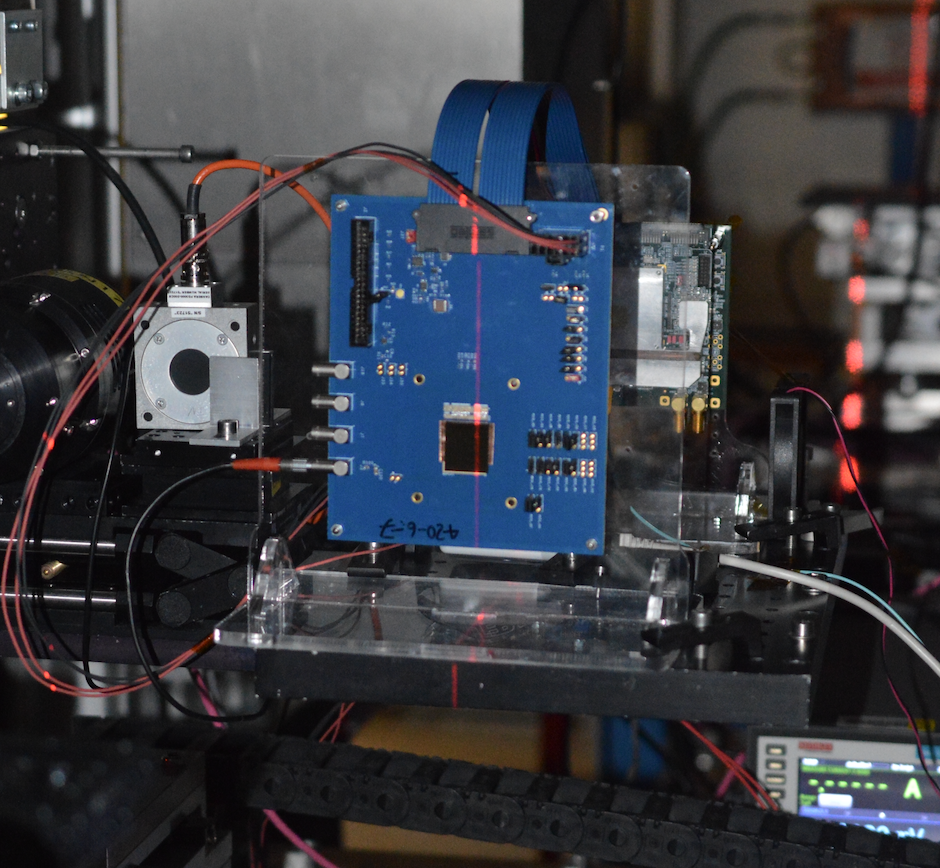}
    \caption{Alignment of the MuPix11 sensor using guidance lasers.}
    \label{fig:alignment}
\end{figure}

The MuPix11 ASIC is mounted on a carrier PCB referred to as the Single Chip Card (SCC). The SCC provides connections to the MuPix11 for low-voltage power, sensor reverse bias, and digital communication. As shown in Fig.~\ref{fig:setup}, the SCC is connected to the Front-End Board (FEB) digital readout system. Data is transmitted from the MuPix11 at \SI{1.25}{Gbps} via LVDS lines to an Arria~V FPGA on the FEB, and subsequently to a Terasic DE5a-Net-DDR4 board inside the DAQ PC using an optical fibre link. This is a simplified version of the final Mu3e readout system \cite{DAQpaper}.
\\
\indent The MuPix11 SCC assembly was installed in the beam hutch of the B16 beamline. To avoid motion-induced smearing during scanning, a mechanical X-ray shutter was integrated into the DAQ and triggered by the motion stage to synchronise data acquisition with stage movement. The chip was approximately aligned with the X-ray beam using guidance lasers, as shown in Fig.~\ref{fig:alignment}.

\subsection{Data Acquisition}
\begin{figure}
    \centering
\includegraphics[width=\linewidth]{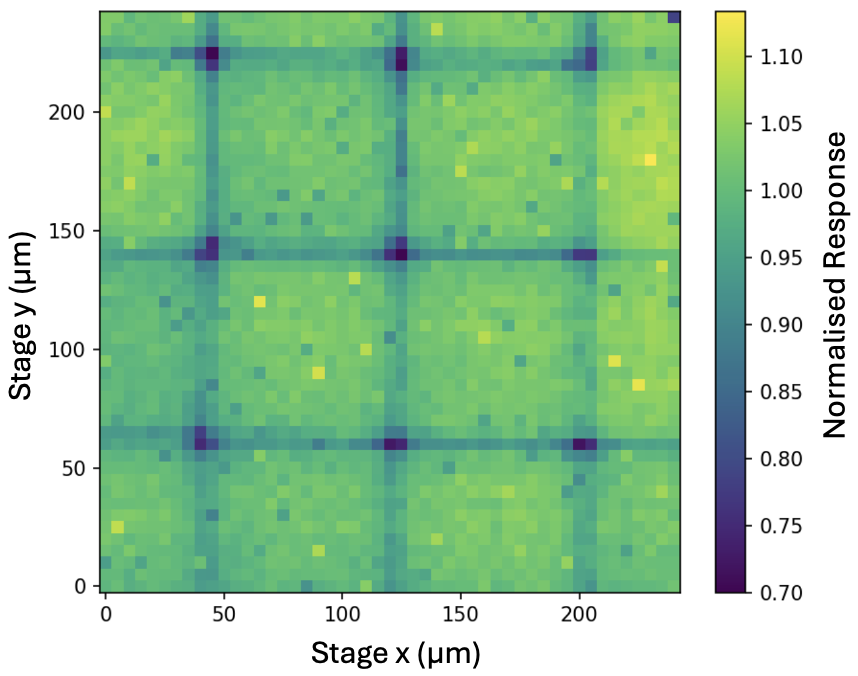}
    \caption{Preliminary two-dimensional scan with \SI{5}{\micro\meter} step size to ensure alignment (bias \SI{0}{V}, threshold \SI{42.2}{\milli\volt}, 6 seconds of beam per point).}
    \label{fig:Scan6}
\end{figure}
The experiment was carried out with a \SI{0.5}{mm} aluminium attenuator in the beam path, yielding a hit frequency of approximately \SI{400}{Hz} for \SI{8}{keV} photons. 
Without attenuation the hit rate increased, but a defocused halo from higher-energy beam harmonics significantly enlarged the micro-focus spot size. Hits were associated with stage coordinates during offline processing by correlating MuPix11 timestamps with trigger timestamps recorded by the beamline DAQ. 

A preliminary two-dimensional scan (Fig.~\ref{fig:Scan6}) was performed with a step size of \SI{5}{\micro\meter}, to determine the beam position relative to the pixel matrix. Due to manual mounting of the MuPix11 ASIC on the SCC and the use of clearance holes for M3 bolts, a small angular misalignment between the pixel matrix and the motion stage was observed.

\section{Results and Analysis}
A summary of the scan configurations is given in Table \ref{tab:scan_summary}. The threshold DAC values 118, 121, and 125 correspond to threshold voltages of \SI{42.2}{\milli\volt}, \SI{63.3}{\milli\volt}, and \SI{91.4}{\milli\volt}, respectively, using a baseline DAC of 112 and a conversion of \SI{7}{\milli\volt}/DAC. These thresholds correspond to approximately 750, 1100, and 1600 electron--hole pairs 
at \SI{-30}{\volt} reverse bias \cite{Lill}.

\begin{table*}
    \caption{Summary of scan configurations.}
    \centering
    \resizebox{\textwidth}{!}{%
    \begin{tabular}{ccccccc}
            \toprule
            Reverse Bias [\SI{}{\volt}] & Threshold [DAC value] & Step Size [\SI{}{\micro\meter}] & Dimensions [\SI{}{\micro\meter}] & Time per Point [\SI{}{\second}] \\
            \midrule
            $0$   & 118 & 5  & 240   & 3  \\
            $-5$  & 118 & 5  & 240   & 3   \\
            $-10$ & 118 & 5  & 240   & 3   \\
            $-20$ & 118 & 5  & 240   & 3   \\
            $-30$ & 118 & 5  & 240   & 3  \\
            $-30$ & 121 & 5  & 240   & 3  \\
            $-30$ & 125 & 5  & 240   & 3  \\
            $0$   & 118 & 3  & 160 $\times$ 160  & 11.7 \\
            $-30$ & 118 & 3  & 160 $\times$ 160  & 15   \\
            \bottomrule
    \end{tabular}
    }
    \label{tab:scan_summary}
\end{table*}

\subsection{2D Scan Results}
\begin{figure}
    \centering
    \includegraphics[width=\linewidth]{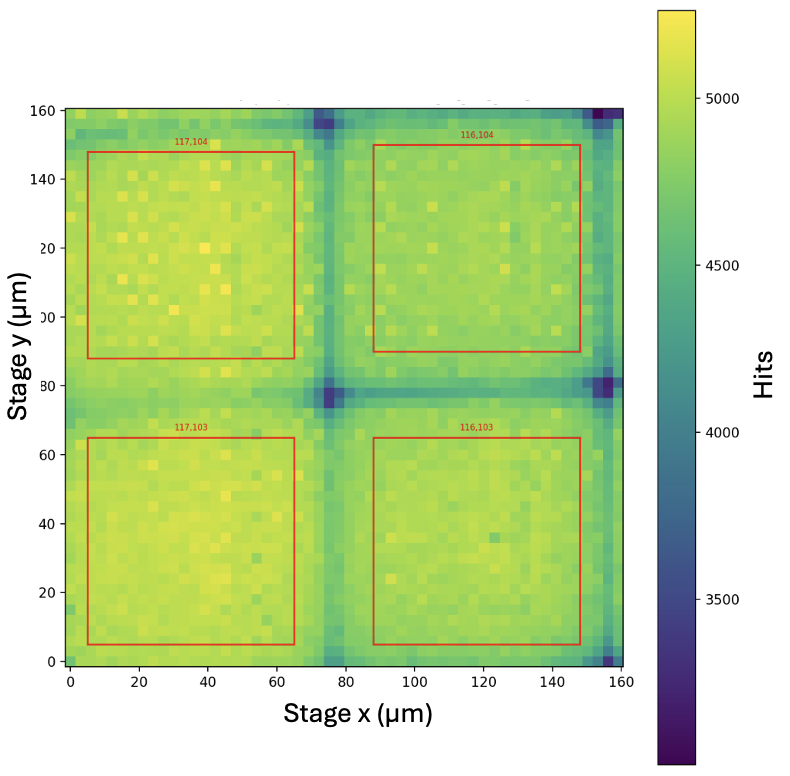}
    \caption{Number of hits recorded at \SI{0}{\volt} bias. Red boxes indicate per-pixel centre averaging regions (\SI{3}{\micro\meter} step size, threshold \SI{42.2}{\milli\volt}, 11.7 seconds of beam per point).}
    \label{fig:0V}
\end{figure}

Two-dimensional scans were performed at \SI{0}{\volt} and the nominal operating voltage of \SI{-30}{\volt} to compare undepleted and fully depleted conditions. The unprocessed hit map at \SI{0}{\volt} (Fig.~\ref{fig:0V}) resolves four pixels, with a visible intensity gradient, likely due to pixel-to-pixel threshold variations.

\begin{figure}
    \centering
    \includegraphics[width=\linewidth]{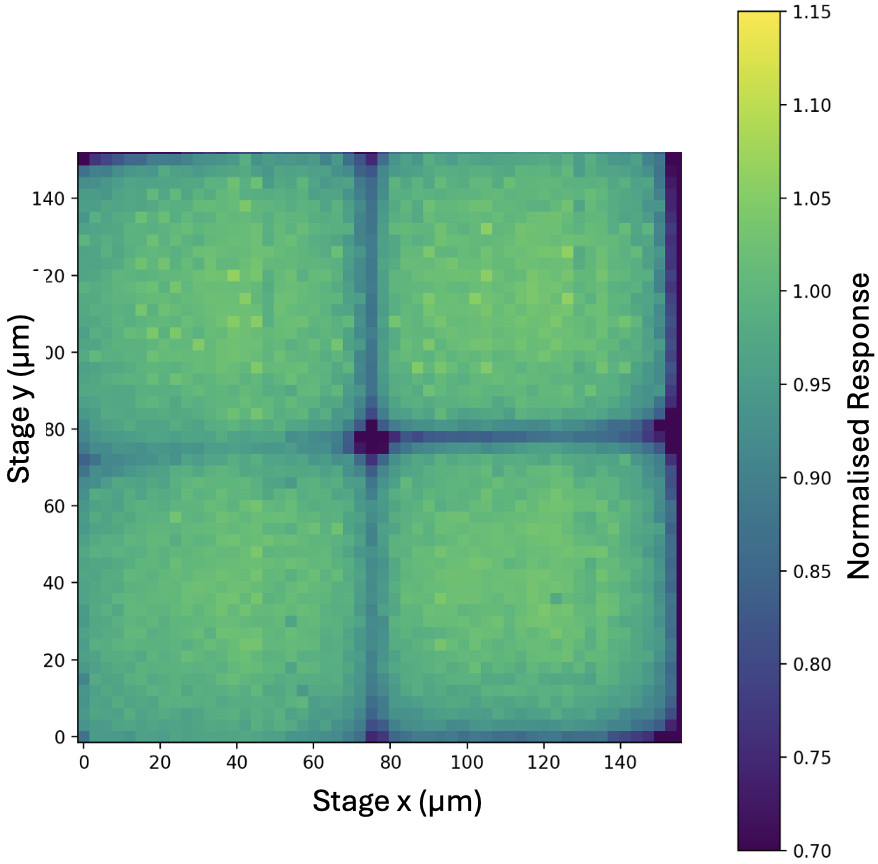}
    \caption{A two-dimensional scan of relative hit rate across four pixels with the sensor biased at \SI{0}{\volt} (\SI{3}{\micro\meter} step size, threshold \SI{42.2}{\milli\volt}, 11.7 seconds of beam per point).}
    \label{fig:0V_averaged}
\end{figure}

\begin{figure*}
    \centering
    \includegraphics[width=\linewidth]{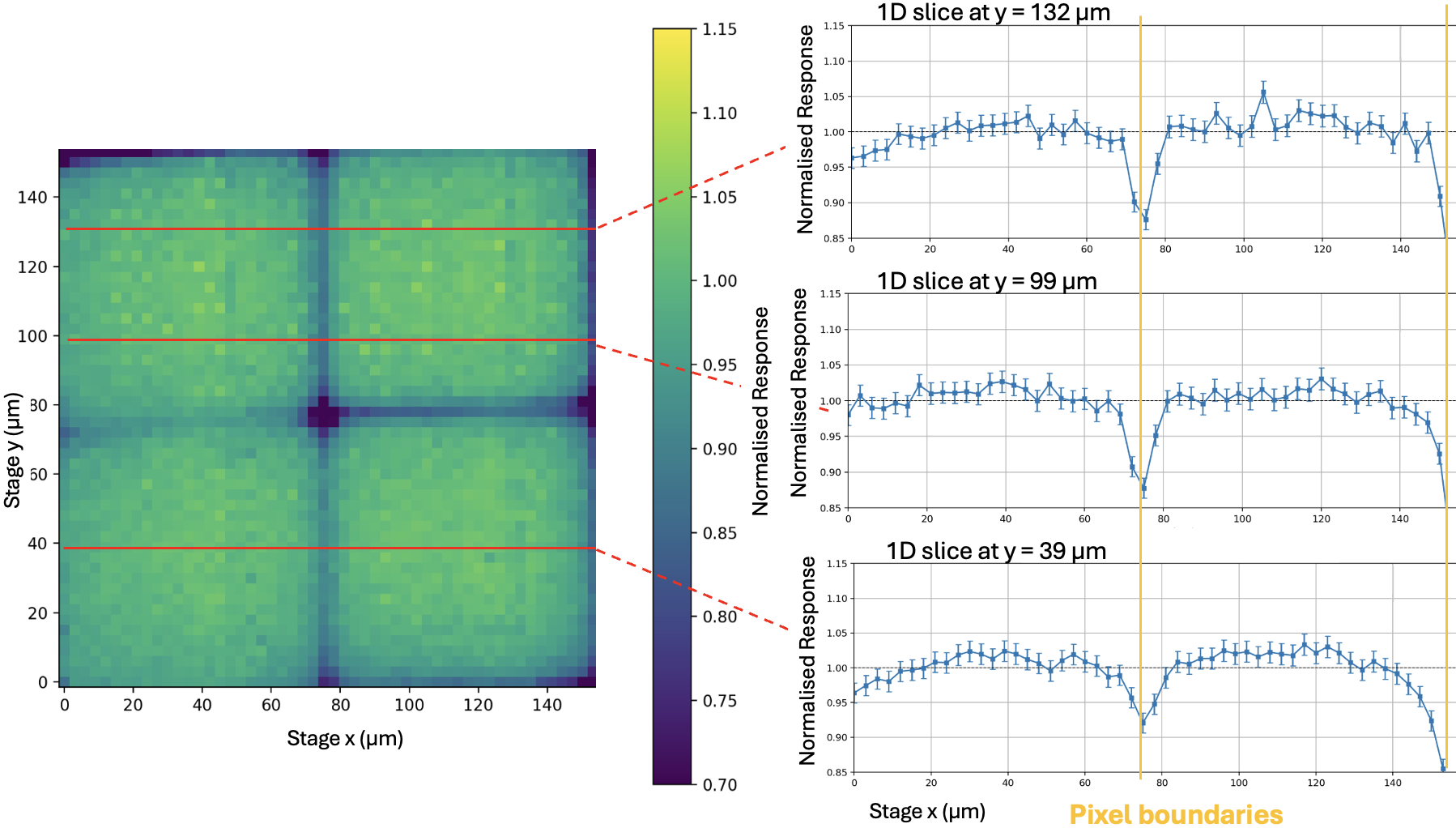}
    \caption{Relative MuPix11 response at \SI{0}{\volt} sensor bias, showing 1D cuts across the pixels at various heights (Poisson error bars). This shows the efficiency drop at pixel boundaries (\SI{3}{\micro\meter} step size, threshold \SI{42.2}{\milli\volt}, 11.7 seconds of beam per point).}
    \label{fig:0V_with_1D}
\end{figure*}

\begin{figure*}
    \centering
    \includegraphics[width=\linewidth]{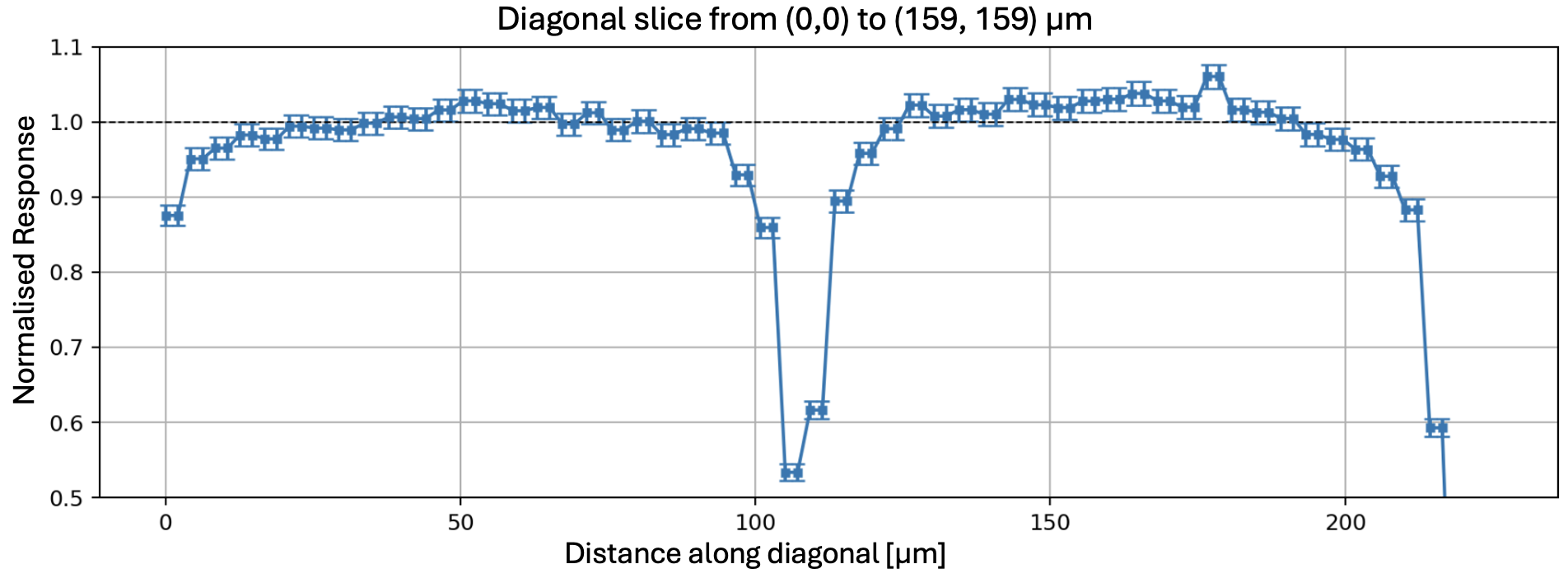}
    \caption{Diagonal cross section of the two-dimensional plot in Fig.~\ref{fig:0V} at \SI{0}{\volt}, showing the hit rate drop at the pixel corner boundary (\SI{3}{\micro\meter} step size, threshold \SI{42.2}{\milli\volt}, 11.7 seconds of beam per point).}
    \label{fig:diagslice}
\end{figure*}

The response is normalised to the mean hit rate at each pixel centre (red boxes). The resulting map (Fig.~\ref{fig:0V_averaged}) shows a $\sim$10\% reduction in hit rate at pixel edges, leading to an effective rounding of the pixel shape. This behaviour is confirmed by the one-dimensional profiles (Fig.~\ref{fig:0V_with_1D}). Along the diagonal (Fig.~\ref{fig:diagslice}), the response at the intersection of four pixels decreases to $\sim$54\%.

\begin{figure}
    \centering
    \includegraphics[width=\linewidth]{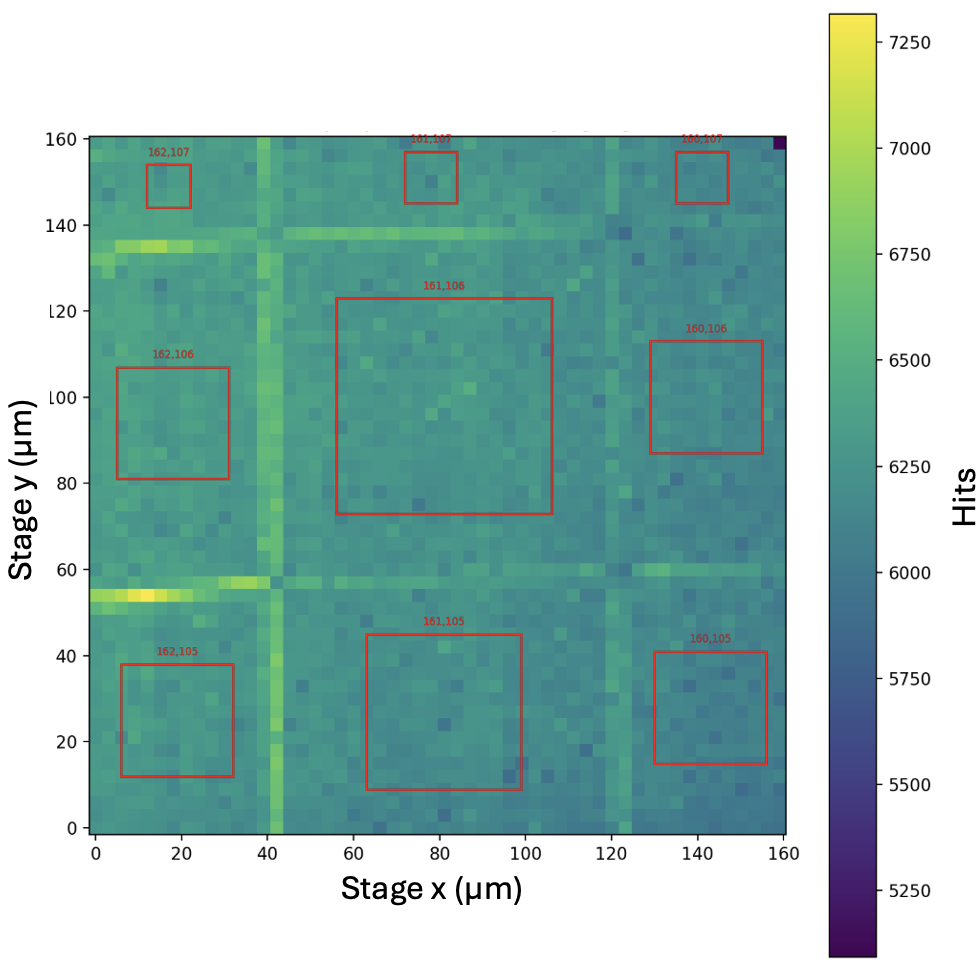}
    \caption{Number of hits recorded at \SI{-30}{\volt} bias. Red boxes indicate per-pixel centre averaging regions (\SI{3}{\micro\meter} step size, threshold \SI{42.2}{\milli\volt}, 15 seconds of beam per location).}
    \label{fig:30V}
\end{figure}

\begin{figure}
    \centering
    \includegraphics[width=\linewidth]{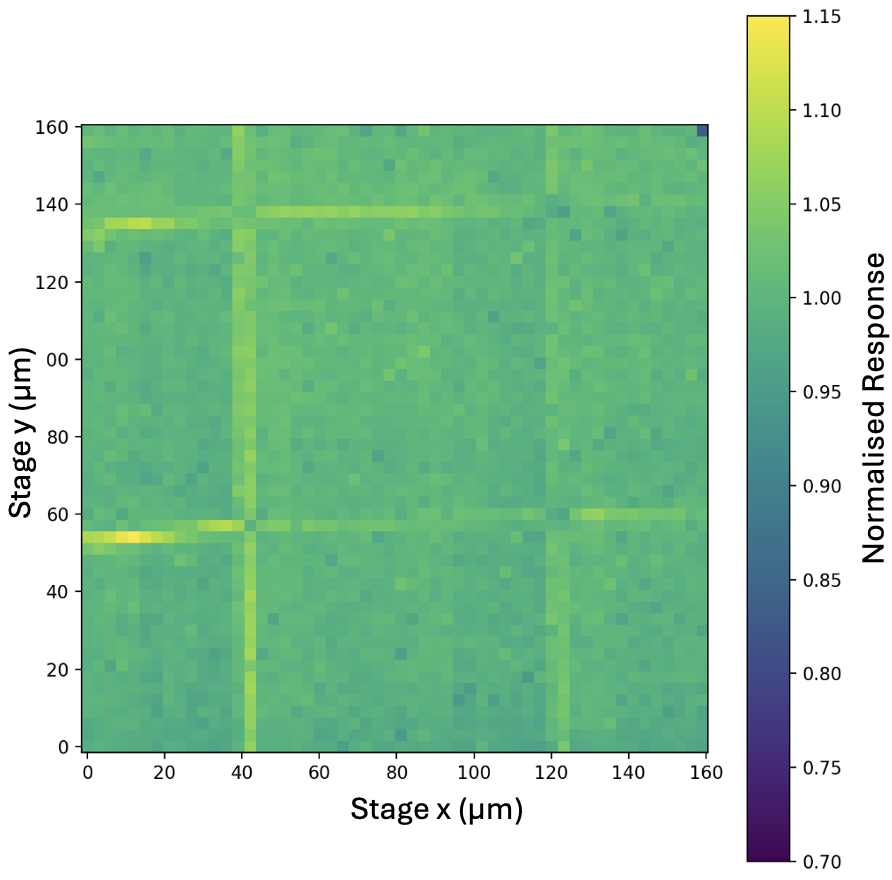}
    \caption{A two-dimensional scan of relative hit rate across a pixel and its neighbouring pixels at \SI{-30}{\volt} sensor bias (\SI{3}{\micro\meter} step size, threshold \SI{42.2}{\milli\volt}, 15 seconds of beam per location).}
    \label{fig:30V_averaged}
\end{figure}

At \SI{-30}{\volt} (Fig.~\ref{fig:30V}), the normalised response (Fig.~\ref{fig:30V_averaged}) is uniform and square, indicating that the depletion region extends across the full pixel area. At pixel boundaries, the response slightly exceeds unity (up to $\sim$1.1), consistent with charge sharing between neighbouring pixels.

\begin{figure*}
    \centering
    \includegraphics[width=\linewidth]{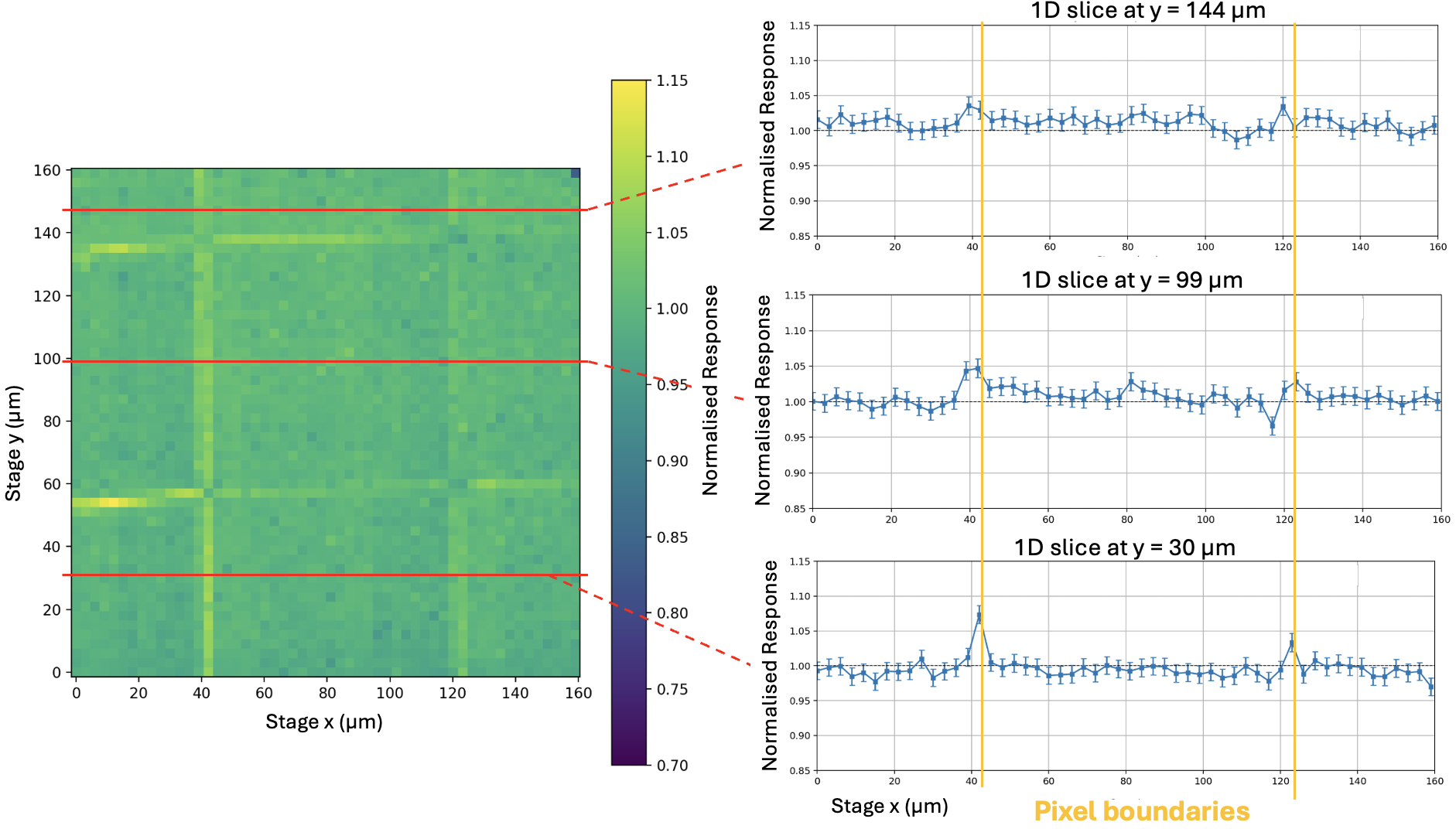}
    \caption{Relative MuPix11 response at \SI{-30}{\volt} sensor bias, with 1D cuts across the pixels at various heights (Poisson error bars) (\SI{3}{\micro\meter} step size, threshold \SI{42.2}{\milli\volt}, 15 seconds of beam per point).}
    \label{fig:30V_with_1D}
\end{figure*}

\begin{figure*}
    \centering
    \includegraphics[width=\linewidth]{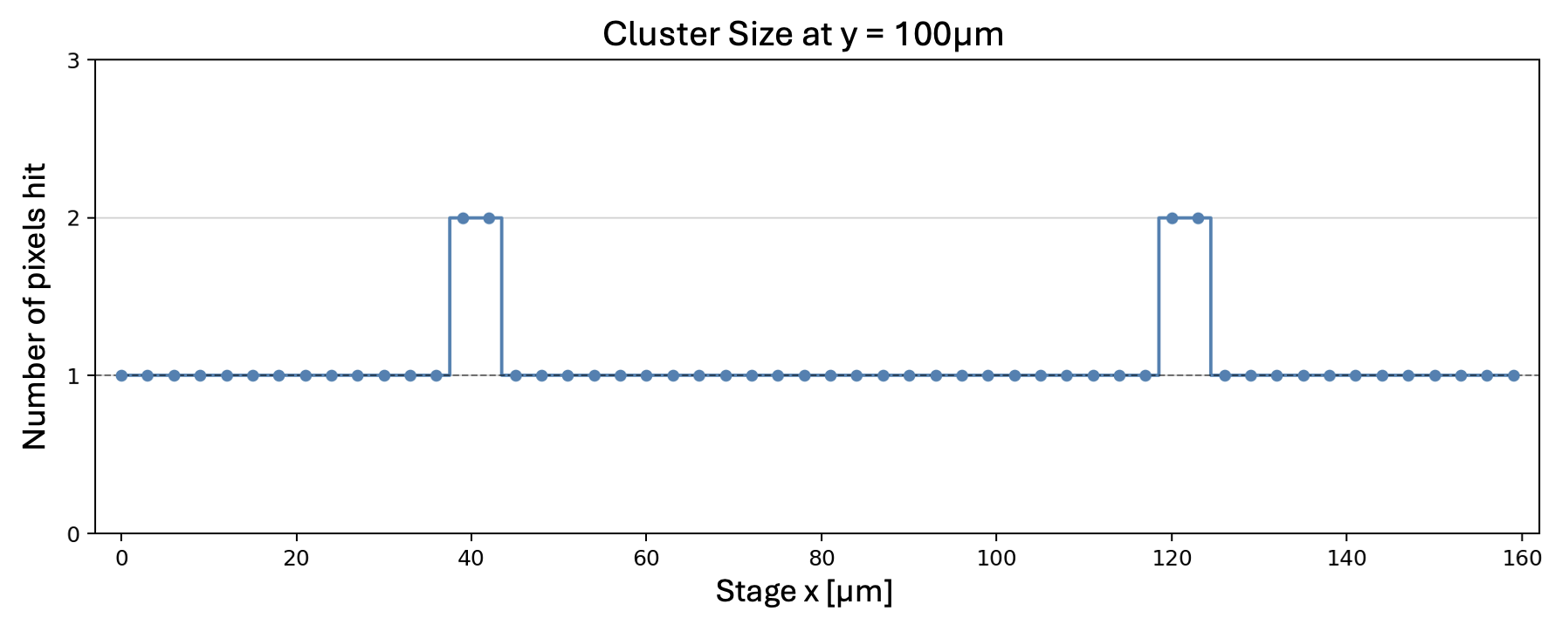}
    \caption{Number of pixels recording a hit along the vertical slice at $y = \SI{100}{\micro\meter}$ at \SI{-30}{\volt}. Pixel boundaries are at \SI{40}{\micro\meter} and \SI{120}{\micro\meter}.}
    \label{fig:clustering}
\end{figure*}

In contrast to the \SI{0}{\volt} case, where incomplete depletion leads to charge loss, full depletion enables charge generated near pixel boundaries to be collected by multiple pixels. This results in cluster formation (Fig.~\ref{fig:clustering}) and an enhanced local hit rate. The one-dimensional profiles (Fig.~\ref{fig:30V_with_1D}) confirm the overall uniformity across the pixel.

\subsection{In-Pixel Efficiency Estimate}
\label{subsec:eff}
Within a pixel, the relative efficiency is estimated at each scan position as the local hit count normalised to the mean hit rate at the pixel centre, as the latter is the region of highest collection efficiency. To avoid double-counting arising from charge sharing with neighbouring pixels, only hits recorded by the pixel under study are considered.

\begin{figure*}
    \centering
    \includegraphics[width=\linewidth]{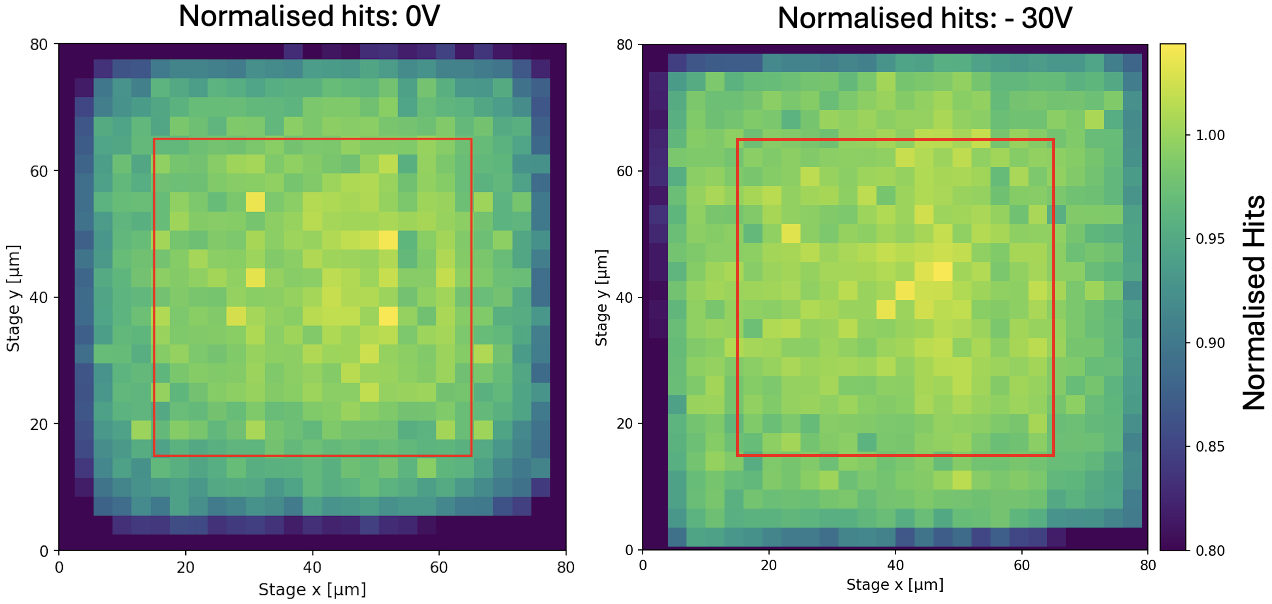}
    \caption{Normalised hits recorded by one pixel inside its physical \SI{80}{\micro\meter} $\times$ \SI{80}{\micro\meter} area: (a) at \SI{0}{\volt}, (b) at \SI{-30}{\volt}. The hits are averaged to the red square in the centre.}
    \label{fig:combined_eff}
\end{figure*}

The normalised hit maps for a single pixel under both bias conditions are shown in Fig.~\ref{fig:combined_eff}. Integrating over the \SI{80}{\micro\meter} $\times$ \SI{80}{\micro\meter} pixel area yields mean normalised responses of $0.957 \pm 0.003$ at \SI{-30}{\volt} and $0.910 \pm 0.004$ at \SI{0}{\volt}. The uncertainties are estimated using the binomial standard error, $\sigma_p = \sqrt{p(1-p)/N}$, with $N$ the total hit count in the reference region.

To quantify the overall performance difference, the efficiency is combined with the measured hit rates ($381 \pm 5$~Hz at \SI{-30}{\volt}, $362 \pm 6$~Hz at \SI{0}{\volt}). The effective throughput is therefore $364.5 \pm 4.6$~Hz and $329.4 \pm 5.8$~Hz, respectively, corresponding to a total reduction of

\begin{equation}
    \Delta = 1 - \frac{\varepsilon_{\text{0V}} \cdot \dot{h}_{\text{0V}}}{\varepsilon_{-\text{30V}} \cdot \dot{h}_{-\text{30V}}} = 9.6 \pm 0.5\%
    \label{eq:eff_loss}
\end{equation}

The hit rate reduction accounts for 5.0\% of this loss and the efficiency drop 
for 4.9\%, with the two effects compounding multiplicatively to give the total 
of 9.6\%.

\subsection{1D Scan Results}

One-dimensional scans spanning \SI{240}{\micro\meter} (three pixel pitches) were performed to study the dependence on bias voltage and threshold. The combined data yield approximately 7200 hits per point.

\begin{figure*}
    \centering
    \includegraphics[width=\linewidth]{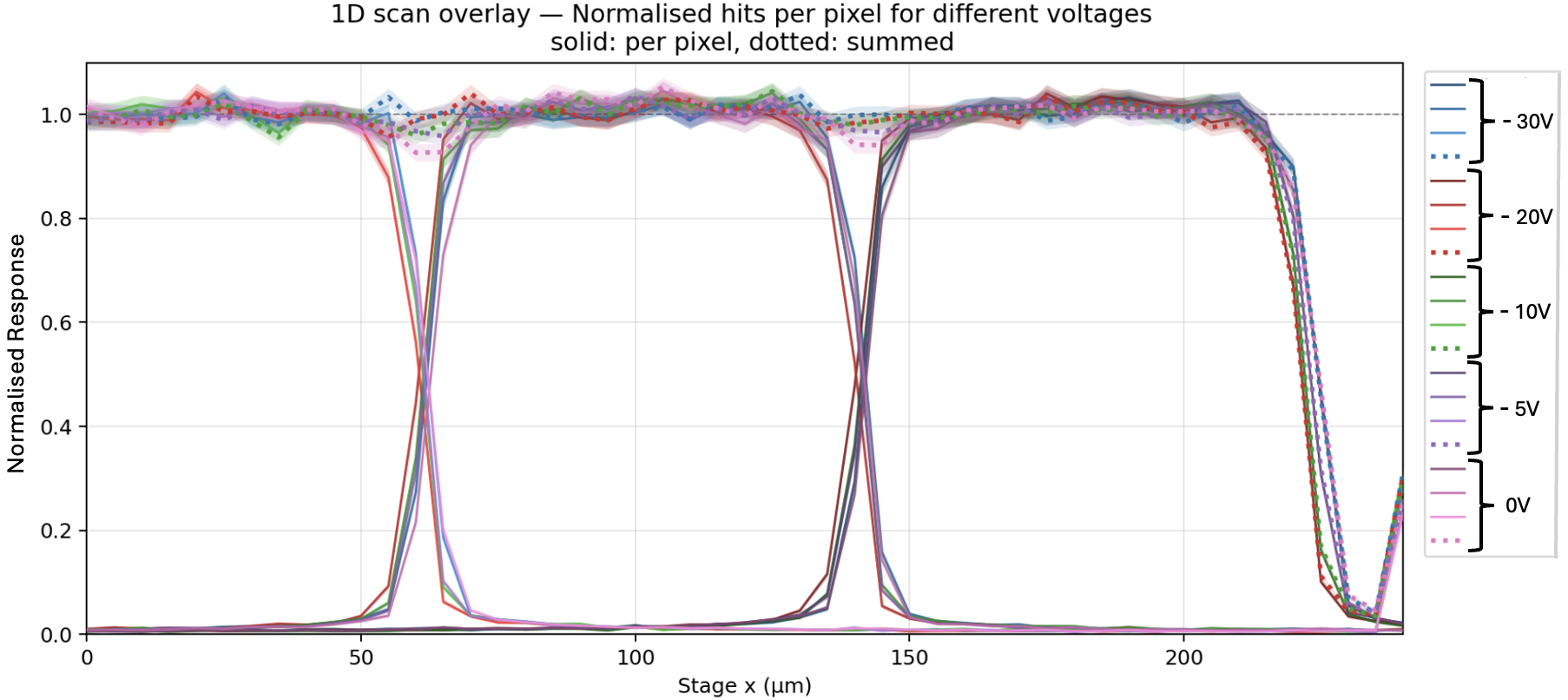}
    \caption{1D scan across three pixels for bias voltages between \SI{0}{V} and \SI{-30}{V}. Each colour indicates a different voltage value. The solid lines correspond to a pixel, while the dotted line is the sum.}
    \label{fig:voltages}
\end{figure*}

\begin{figure*}
    \centering
    \includegraphics[width=\linewidth]{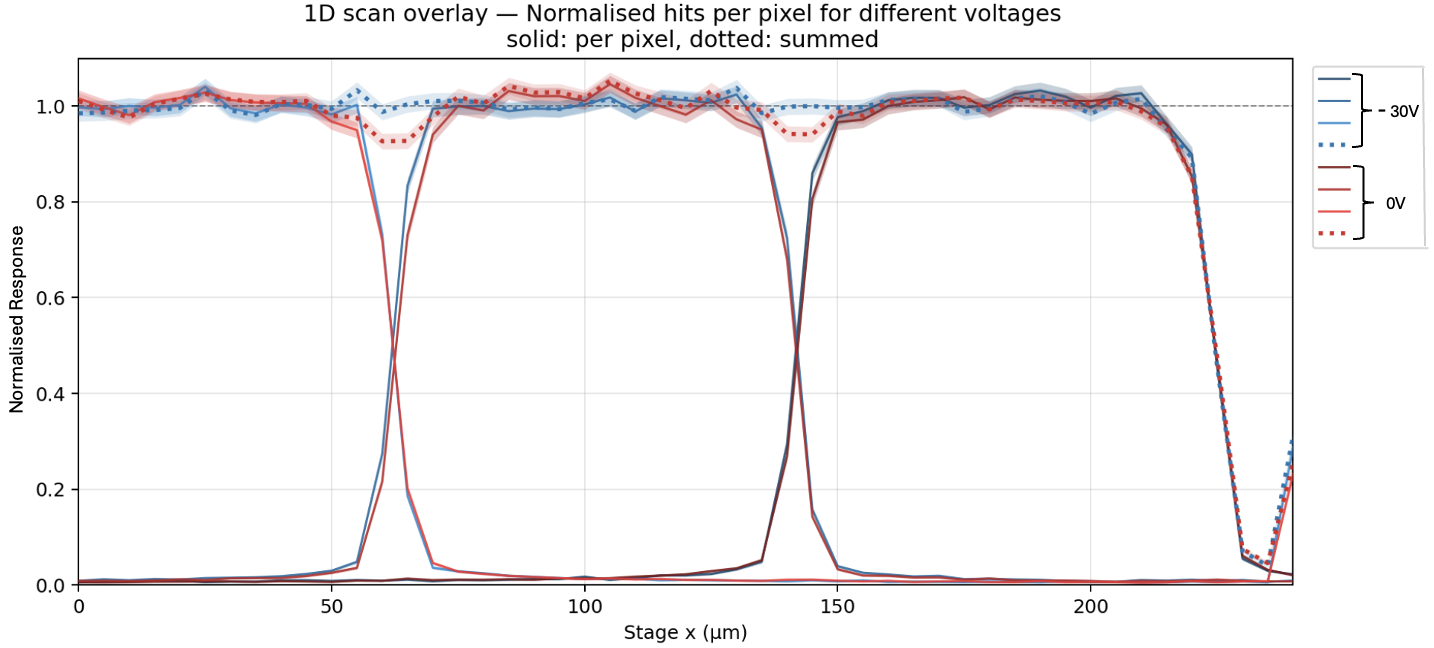}
    \caption{1D scan across three pixels at different bias voltages. Intermediate scans between \SI{0}{V} and \SI{-30}{V} removed for clarity. Each colour indicates a different voltage value. The solid lines correspond to a pixel, while the dotted line is the sum.}
    \label{fig:voltages_reduced}
\end{figure*}

Figure~\ref{fig:voltages} shows scans for bias voltages between \SI{0}{\volt} and \SI{-30}{\volt}. A reduction in response at pixel edges is observed with decreasing bias, consistent with the 2D measurements. Figure~\ref{fig:voltages_reduced} highlights the comparison between \SI{0}{\volt} and \SI{-30}{\volt}.

\begin{figure*}
    \centering
    \includegraphics[width=\linewidth]{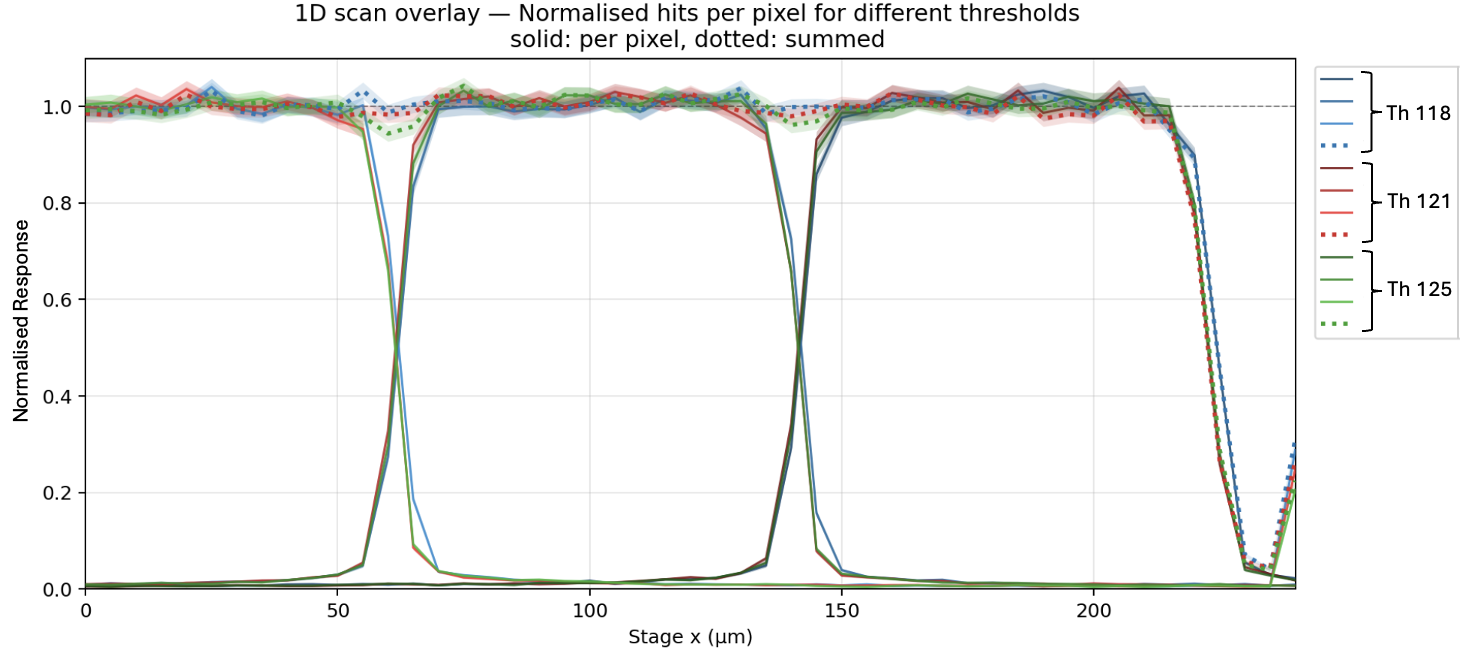}
    \caption{1D scan across three pixels for thresholds between \SI{42.2}{mV} -- \SI{91.4}{mV}. Each colour indicates a different threshold value. The solid lines correspond to a pixel, while the dotted line is the sum.} 
    \label{fig:threshold}
\end{figure*}

The threshold dependence (Fig.~\ref{fig:threshold}) is weak over the range studied (DAC 118–125), with slightly higher efficiency at lower threshold. However, the variation is comparable to statistical uncertainties.


\subsection{Discussion}

The pixel centres exhibit similar hit rates in both the zero bias and operational configurations, differing only by 3.7\%. This indicates that, for MuPix11, the absence of external bias does not preclude signal detection. As discussed in Section \ref{subsec:HVMAPS}, the built-in potential of the p–n junction and the n-well bias create a localised depletion region around the collection n-well. Consequently, charge generated by the X-rays near the collection electrode is efficiently collected by drift. In addition, a fraction of the charge carriers generated further away can reach the electrode via diffusion within the shaping time of the charge-sensitive amplifier.
\\
\indent In contrast, a clear efficiency loss is observed at pixel boundaries at \SI{0}{\volt}.
Here photons are not converting directly above the collection electrode. In the zero bias case, the depletion region does not reach the pixel edges and charge transport is therefore dominated by diffusion. The observed efficiency reduction is attributed to carrier recombination during transport or diffusion into neighbouring pixels. The magnitude of this effect remains moderate due to the sensor thickness of \SI{70}{\micro\meter}, which limits the distance to the collection electrode. Furthermore, an \SI{8}{keV} X-ray produces approximately 2,200 electron-hole pairs, such that even partial charge loss often still results in a signal above threshold.
\\
\indent A systematic pattern is observed across all pixels, appearing as a vertical stripe in the right half of the sensor. This is more pronounced at \SI{0}{V}, but remains visible at \SI{-30}{V}. As the sensor is being illuminated from the processed side, this structure is likely associated with a metal routing line within the chip.

\section{Conclusion}

The sub-pixel response of a MuPix11 sensor was mapped using a focused \SI{8}{keV} X-ray beam at the B16 beamline of Diamond Light Source. At the nominal operating voltage, a uniform response is observed across the pixel area, indicating that the depletion region extends laterally over the full pixel cell, with charge sharing at the boundaries.

The pixel centre exhibits comparable hit rates at \SI{0}{\volt} and \SI{-30}{\volt}, consistent with a localised depletion region formed by the built-in p--n junction potential. This enables efficient drift-based charge collection near the electrode even in the absence of external bias.



\section*{Acknowledgments}
We acknowledge Diamond Light Source for time on beamline B16 under Proposal OM41168-2, and extend our thanks to beamline scientists V. Dhamgaye and I. Dolbnya. We also thank M. Larner-Smith who built custom supports for PCB and FEB.  
  
\subsection*{Funding}
This work was supported by the Science and Technology Research Council [grant number ST/W000628/1]; Beamtime and support awarded by Diamond Light Source; the Mu3e Experimental Collaboration and the Clarendon Scholarship.

\subsection*{Conflicts of Interest}
The authors declare that there is no conflict of interest regarding the work reported in this paper.

\bibliographystyle{unsrtnat}
\bibliography{DLS2026_new}

\end{document}